\documentclass[11pt,a4paper]{article} 
\usepackage{jheppub,hyperref,mdwlist}
\usepackage{jheppub,hyperref,mdwlist}
\usepackage{amsmath}
\usepackage{graphicx}
\usepackage{subfigure}
\usepackage{amssymb}
\usepackage{xcolor}
\usepackage{multirow}
\usepackage{cancel}
\usepackage{color}
\usepackage{mathtools}
\usepackage{listings}
\usepackage{xcolor}
\usepackage{float}
\usepackage{slashed}
\usepackage{bm}
\usepackage{amsmath}
\usepackage{wrapfig}

\newcommand{\be}{\begin{equation}}
\newcommand{\ee}{\end{equation}}
\newcommand{\beq}{\begin{equation}}
\newcommand{\eeq}{\end{equation}}
\newcommand{\bea}{\begin{eqnarray}}
\newcommand{\eea}{\end{eqnarray}}
\definecolor{airforceblue}{rgb}{0.36, 0.54, 0.66}
\definecolor{steelblue}{rgb}{0.27, 0.51, 0.71}
\definecolor{amber}{rgb}{1.0, 0.49, 0.0}

\title{  Shedding light on the electroweak phase transition from exotic Higgs boson decays at the lifetime frontiers}

\author[a]{Wei Liu}
\affiliation[a]{Department of Applied Physics, Nanjing University of Science and Technology, \\
Nanjing, 210094, P.R.China}
\author[b]{Aigeng Yang}
\affiliation[b]{Institute of Theoretical Physics, School of Physics, Dalian University of Technology, No.2 Linggong Road, Dalian, Liaoning, 116024, P.R.China }
\author[b]{Hao Sun}
\emailAdd{wei.liu@njust.edu.cn}
\emailAdd{aigengyang@mail.dlut.edu.cn}
\emailAdd{haosun@dlut.edu.cn}

\abstract{
We study the scenarios where a strongly first-order electroweak phase transition (EWPT) is triggered by a light singlet scalar, which has feeble interactions to the Higgs. 
Since the singlet scalar is light and has weak couplings, it can decay at a macroscopic distance away from the collision point. Therefore, it can be regarded as a long-lived particles (LLP) in such scenarios. We perform the searches of the LLPs from the exotic Higgs decays, at the FASER, MAPP and CMS-Timing detectors of the 14 TeV HL-LHC, to probe the strongly first-order EWPT. In certain scenarios, we show that the LLP searches can help to reach the parameter space of the strongly first-order EWPT remarkably, where the searches for promptly exotic Higgs decays are not valid. }

\begin{document}
\maketitle
\flushbottom

\section{Introduction}
\label{sec:intro}
After the discovery of the Higgs boson at the LHC~\cite{ATLAS:2012yve, CMS:2012qbp}, the pattern of the electroweak phase transition~(EWPT) is still unknown,
remained as one of the most important problems of the particle physics. Although the Standard Model~(SM) predicts a smooth crossover of the EWPT in the early universe~\cite{Kajantie:1996qd,Rummukainen:1998as,Laine:1998jb}, there is another intriguing possibility that the EWPT can be strongly first-order~(SFOEWPT). A SFOEWPT is indeed interesting, as it can accommodate the observation of the baryon asymmetry of the Universe via the electroweak baryogenesis mechanism~(EWBG)~\cite{Cline:2006ts, Trodden:1998ym, Morrissey:2012db}.

In general, there are two ways to probe the EWPT experimentally, either via the phenomenology of the Higgs related physics at the colliders, or the primordial stochastic gravitational waves~(GW) generated during the EWPT at the GW detectors~\cite{Caprini:2015zlo,Caprini:2019egz,Cepeda:2019klc,DiMicco:2019ngk,FCC:2018byv}. Many studies have been done, either via focusing on one of the phenomena or comparing the both, based at different new physics models beyond the SM,
including the real singlet extended SM ~\cite{McDonald:1993ey,Profumo:2007wc,Espinosa:2011ax,Cline:2012hg,Alanne:2014bra,Profumo:2014opa,Alves:2018jsw,Vaskonen:2016yiu,Huang:2018aja,Cheng:2018ajh,Alanne:2019bsm,Gould:2019qek,Carena:2019une,Ghorbani:2018yfr,Ghorbani:2017jls,Liu:2021jyc, Carena:2022yvx,Blasi:2022woz,Chen:2017qcz}, two-Higgs-doublet model~\cite{Turok:1990zg,Turok:1991uc,Cline:2011mm,Dorsch:2013wja,Chao:2015uoa,Basler:2016obg,Haarr:2016qzq,Dorsch:2017nza,Andersen:2017ika,Bernon:2017jgv,Wang:2018hnw,Wang:2019pet,Kainulainen:2019kyp,Su:2020pjw}, left-right symmetric model~\cite{Brdar:2019fur,Li:2020eun}, Georgi-Machacek model~\cite{Zhou:2018zli}, composite Higgs models~\cite{Espinosa:2011eu,Chala:2016ykx,Chala:2018opy,Bruggisser:2018mus,Bruggisser:2018mrt,Bian:2019kmg,DeCurtis:2019rxl,Xie:2020bkl,Xie:2020wzn}, as well as SUSY models~\cite{Wang:2022lxn}, etc.
Among these studies, Ref.~\cite{Liu:2021jyc} investigates the complementarity between the proposed multi-TeV muon colliders and the LISA GW detectors to the first-order EWPT based at the real singlet extended SM. 
In most cases, the mass of the BSM particles responsible for the interactions driving a SFOEWPT is taken to be larger than the electroweak scale.

As motivated by Ref.~\cite{Profumo:2007wc,Kozaczuk:2019pet,Carena:2019une,Davoudiasl:2021vku, Carena:2022yvx}, a light BSM particle with masses below the electroweak scale can also trigger a SFOEWPT. If its mass is below $m_Z$, it must be a singlet-like scalar~(we denote the mass eigenstate as $h_1$) as argued by Ref.~\cite{Kozaczuk:2019pet}. The direct production of such scalar is controlled by the Higgs mixings~$\theta$, which is tiny at the current limits via the Higgs measurements and exotic Higgs seaches at the LHC and LEP~\cite{Robens:2015gla}, such as $\theta \lesssim 10^{-2}$. Nonetheless, a first-order EWPT scenario can lead to appreciable direct couplings between the singlet-like scalar and SM Higgs, results in sizeable $h_2 \rightarrow h_1 h_1$ branching ratio, where the $h_2$ is the mass eigenstate of the observed Higgs. Therefore, promptly exotic Higgs decay searches at the current LHC as well as the future Higgs machines can be used to probe the EWPT as summarised in Ref.~\cite{Carena:2022yvx}. There are still plenty of parameters open for the SFOEWPT, and the promptly exotic Higgs decay searches exclude the scenarios where $m_{h_2} \gtrsim 30$ GeV at the current LHC.

In search of additional sensitivity, one can consider another possible signature of their final states. Searches for the long-lived particles~(LLP) is a hot topic in recent years. Specified detectors at the lifetime frontier for this purpose are proposed and designed, in which the FASER~\cite{Feng:2017uoz} and MoEDAL-MAPP~(MAPP)~\cite{Pinfold:2019nqj}, as well as the precision timing detector at the CMS~(CMS-Timing)~\cite{Butler:2019rpu, Liu:2018wte} are already in installtion.
Benefited from the low SM background, they can probe new physics with high sensitivity. Studies about the possible candidates of the LLPs at the LHC is widely discussed, including the axion, heavy neutral lepton, dark photon as well as the light scalar, etc~(see Ref.~\cite{Alimena:2019zri} and references therein). Ref.~\cite{Deppisch:2018eth,Deppisch:2019kvs,Liu:2022kid,Liu:2022ugx} treated the heavy neutrinos as the LLP candidate, and performed the detailed analyses of the LLP searches to test the type-I seesaw mechanisms.
If the Higgs mixings are much smaller than the current limits, i.e. $\theta \lesssim 10^{-4}$, or the singlet-like scalar is light, it can decay at a macroscopic distance away from the collision point. Hence, the EWPT can also be probed by the searches of the LLPs. 
A study about the long-lived light singlet scalar via the CMS phase II track trigger has already been done at Ref.~\cite{Gershtein:2020mwi}.
Due to the low SM background, the LLP searches can potentially be sensitive to much lower branching ratio of the $h_2 \rightarrow h_1 h_1$, probing much more cases of the EWPT.

This possibility is however rarely been investigated. We therefore focus on the searches of the exotic Higgs decays at the lifetime frontiers, and compare it to the LISA GW detector in probing the EWPT. As the real singlet extended SM is one of the simplest models which can introduce the SFOEWPT, we take it as our benchmark model. The most important features of the SFOEWPT induced by tree level barrier via renormalizable operators~\cite{Chung:2012vg}, is already captured by this model. We perform the studies of the collider phenomenology at the 14 TeV HL-LHC, with an intergated luminosity of 3 ab$^{-1}$, and compare the sensitivity from the searches of the LLP to those from the promptly exotic Higgs decays as summarised in Ref.~\cite{Carena:2022yvx}.

This paper is organized in following order. In Section~\ref{sec:xSM}, we introduce the real singlet extended SM and the corresponding parameter space for SFOEWPT, as well as its GWs signals at the LISA detector. In Section~\ref{sec:LLP}, the phenomenology of the long-lived singlet-like scalar is studied, including its production cross section and decay length, as well as the detailed LLP analyses. The sensitivity to the EWPT is shown in the Section~\ref{sec:sen}, where the projected limits from the promptly exotic Higgs decays at the HL-LHC is also discussed for comparison. In the end, we draw our conclusion in Section~\ref{sec:conclu}.

\section{FOEWPT triggered by a light singlet scalar}
\label{sec:xSM}
\subsection{The real singlet extended SM}
In tree level, the scalar potential of the real singlet extended SM can be expressed as
\be\label{V}
V=-\mu^2|H|^2+\lambda|H|^4+\frac{a_1}{2}|H|^2S+\frac{a_2}{2}|H|^2S^2+b_1S+\frac{b_2}{2}S^2+\frac{b_3}{3}S^3+\frac{b_4}{4}S^4,
\ee
where $H$ and $S$ are the scalar fields for the SM Higgs doublet and singlet, respectively. There are eight parameters, while the shift invariance of the singlet potential removes one degree of freedom as we take $b_1 = 0$, and the measurements of the Higgs mass and electroweak vacuum expectation value removes two. Therefore, we have five free parameters.

After the electroweak symmetry breaking, we can expand the fields around the vacuum in unitary gauge,
\be
H=\frac{1}{\sqrt2}\begin{pmatrix}0\\ v+h\end{pmatrix},\quad S=v_s+s,
\ee
where $v=$ 246 GeV is the vacuum expectation value~(VEV) of the electroweak, and $v_s$ is the VEV for the singlet. The physical scalar masses can be obtained via diagonalizing the Hessian matrix at the physical VEV,
\be
\mathcal{M}_s^2=\begin{pmatrix}\frac{\partial^2V}{\partial s^2}&\frac{\partial^2V}{\partial s\partial h}\\ \frac{\partial^2V}{\partial s\partial s}& \frac{\partial^2V}{\partial h^2}\end{pmatrix}.
\ee
Without additional symmetries, the two mass eigenstates are mixed as
\be
\label{eq:trans}
\begin{split}
h_1 =& s \cos \theta + h \sin \theta,\\
h_2 =& -s \sin \theta + h \cos \theta,
\end{split}
\ee
where we choose $M_{h_1} < M_{h_2} = M_{h}^{\text{SM}} = 125.09$ GeV. We are more interested in the low mixing cases, where the $h_1$ is  mostly singlet-like, and the $h_2$ is the observed Higgs. 
Note that we have defined the mixing angle $\theta$ inversely comparing to Ref.~\cite{Kozaczuk:2019pet}.

As these parameters are directly linked to the experimental observable, we use $M_{h_1}$ and $\theta$ as input free parameters. The coefficients $\mu, b_2, \lambda, a_1$ and $v_s$ can be expressed as 

\be\label{e2}\begin{split}
\mu^2=& \lambda v^2+\frac{v_s}{2}(a_1+a_2v_s),\\ b_2=&-\frac{1}{4v_s}\left[v^2(a_1+2a_2v_s)+4v_s^2(b_3+b_4v_s)\right],\\
\lambda=&~\frac{M_{h_1}^2s_\theta^2+M_{h_2}^2c_\theta^2}{2v^2},\\
a_1=&~\frac{4v_s}{v^2}\left[v_s^2\left(2b_4+\frac{b_3}{v_s}\right)-M_{h_1}^2c_\theta^2-M_{h_2}^2s_\theta^2\right],\\
v_s=&~\frac{1}{2a_2}\left[\frac{s_{2\theta}}{v}\left(M_{h_1}^2-M_{h_2}^2\right)-a_1\right],
\end{split}\ee
where $s_\theta$ and $c_\theta$ are in short of $\sin \theta$ and $\cos \theta$, respectively.

Therefore, we choose the following five free parameters, 
\be\label{input}
\left\{M_{h_1},\theta,a_2,b_3,b_4\right\}.
\ee

\subsection{Strongly first order electroweak phase transition in the small-mixing limit}

The scalar potential in the early universe receives thermal corrections, so Eq.~\ref{V} becomes
\be
V_{\text{eff}} = V + V_T,
\ee
where
\be\label{VT}\begin{split}
V_T=&-\left(\mu^2-c_HT^2\right)|H|^2+\lambda|H|^4+\frac{a_1}{2}|H|^2S+\frac{a_2}{2}|H|^2S^2\\
&+\left(m_1T^2\right)S+\frac{b_2+c_ST^2}{2}S^2+\frac{b_3}{3}S^3+\frac{b_4}{4}S^4,
\end{split}\ee
in the finite temperature~\cite{Liu:2021jyc}, we only keep the $\mathcal{O} (T^2)$ terms~\cite{Profumo:2007wc,Profumo:2014opa}, and the coefficients are
\be
c_H=\frac{3g^2+g'^2}{16}+\frac{y_t^2}{4}+\frac{\lambda}{2}+\frac{a_2}{24},\quad
c_S=\frac{a_2}{6}+\frac{b_4}{4},\quad
m_1=\frac{a_1+b_3}{12}.
\ee
According to Ref.~\cite{Liu:2021jyc}, the tadpole term for the $s$ should be kept since it has considerable impact on the EWPT pattern.

The vacuum structure of the scalar potential is modified by the thermal corrections, especially at high temperatures. We define the critical temperate $T_c$, at which there exists two degenerate vacua.
At temperature below $T_c$, the EW-broken vacuum is more stable, and the Universe decay to it spontaneously, with the decay rate per unit volume as~\cite{Linde:1981zj}
\be
\Gamma(T)\sim T^4\left(\frac{S_3(T)}{2\pi T}\right)^{3/2}e^{-S_3(T)/T}.
\ee
Here, $S_3(T)$ is the Euclidean action of the $O(3)-$ symmetric bounce solution. First-order EWPT happens when the decay is strong in the expanding Universe, i.e. $\Gamma(T)/H^4(T) \gtrsim \mathcal{O}(1)$, where $H(T)$ is the Hubble constant at temperature $T$. The nucleation temperature $T_n$ when the EW-broken vacuum just starts to nucleate can be defined as $\Gamma(T_n) = H^4(T_n)$, and it can be solved approximately using~\cite{Quiros:1999jp}
\be\label{FOPT}
S_3(T_n)/T_n\approx140,
\ee
in the radiation-dominated Universe with phase transition at EW scale. 
To accommodate successful EWBG, the phase transition is required to be strong, i.e. SFOEWPT,
\be
v^{f}/T_n \gtrsim 1,
\ee
where $v^{f}$ is the final VEV of the electroweak after phase transition.

Before we carry on the numerically scan to generate the data samples in the parameter space where SFOEWPT can happen, we noted that there exists semi-analytical bounds on them as described in Ref.~\cite{Kozaczuk:2019pet}. This is realised by approximately adopting a $Z_2$ symmetry for the $S$ field, when the mixing is small~\footnote{The $Z_2$ symmetry might be broken, so the general expression in Eq.~\ref{V} is still valid.}. In this situation, the EWPT pattern can be treated approximately as
\be\begin{split}
(h=0, s \simeq 0) \rightarrow (h=0, s \neq 0) \rightarrow (h\neq 0, s \simeq 0),
\end{split}\ee
so called two-step transitions. 

\begin{figure}[htbp]
\centering
  \includegraphics[scale=0.45]{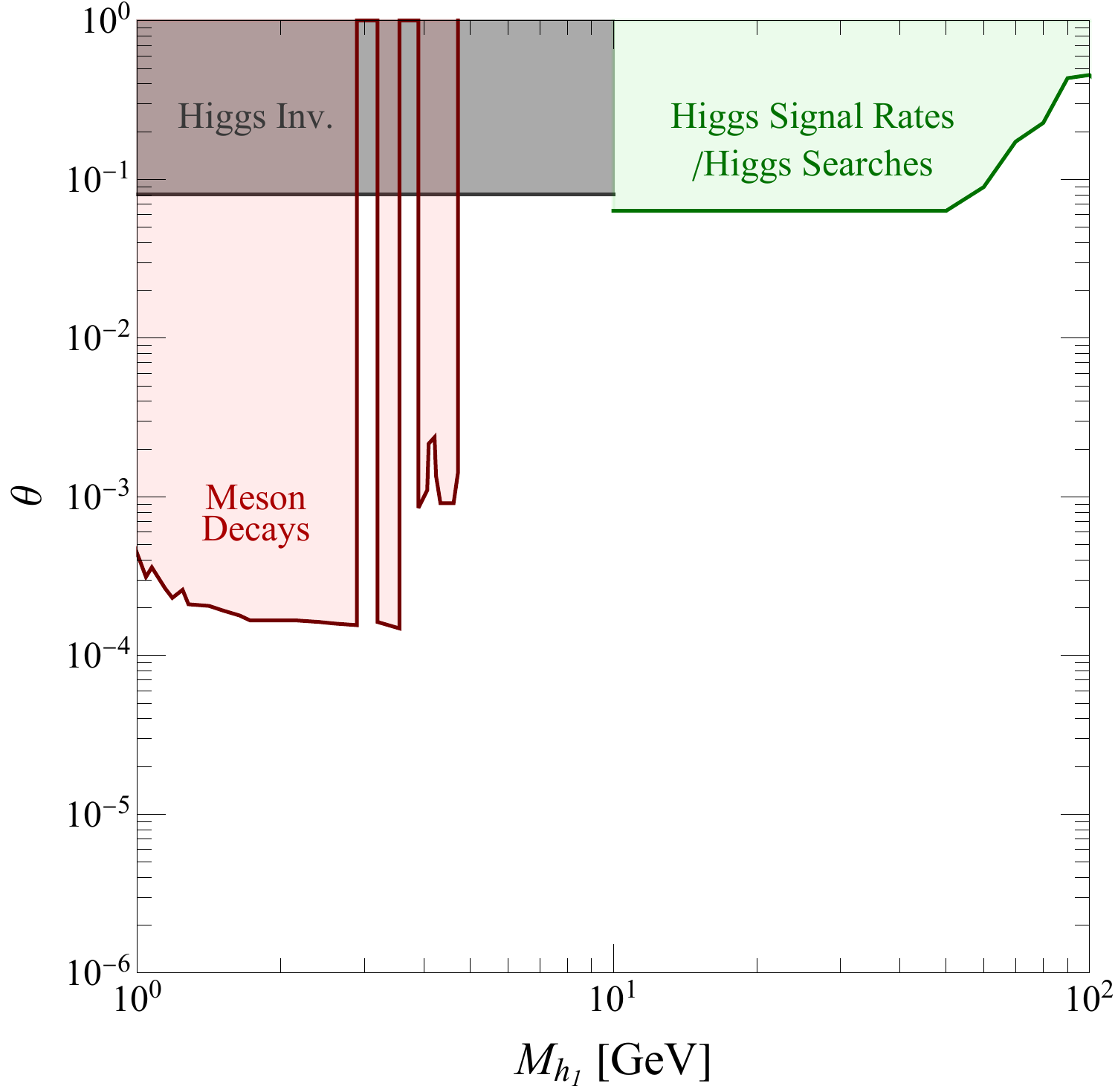}
  \caption{The current limits on the ($M_{h_1}$,$\theta$) from the measurements of the Higgs signal rates and Higgs searches at the LEP and LHC~\cite{Bechtle:2013xfa,Bechtle:2013wla,Robens:2015gla}, the measurements of the Higgs invisible decays~\cite{ATLAS:2020kdi,Dev:2020jkh,Dev:2021kje},
  as well as the measurements of rare meson decays at the LHCb~\cite{LHCb:2016awg}.}
\label{Fig:sa}
\end{figure}

The mixing $\theta$ is well-constrained, see Fig.~\ref{Fig:sa} for the current limits on $(M_{h_1},\theta)$ plane. For $M_{h_1} \gtrsim$ 10 GeV,
the measurements of the Higgs signal rates and light Higgs searches at the LEP and LHC~\cite{Bechtle:2013xfa,Bechtle:2013wla, Robens:2015gla} put stringent limits on the mixing. The measurements of the Higgs invisible decays put bound $\theta \lesssim 0.079$ from $\text{Br}_{h_2 \rightarrow \text{inv}}\lesssim 0.11$ when $m_{h_1} \ll m_{h_2}$~\cite{ATLAS:2020kdi,Dev:2020jkh,Dev:2021kje}.
When the $h_1$ is even lighter than mesons, the measurements of rare meson decays at the LHCb can be applied~\cite{LHCb:2016awg}. Therefore, from the figure, in most of the parameter space, we have $\theta \lesssim \mathcal{O} (10^{-1})$, especially for $M_{h_1} < M_{h_1}/2 \approx $ 62.5~GeV where the exotic decays of the Higgs $h_2 \rightarrow h_1 h_1$ can happen. The HL-LHC with much larger integrated luminosity, and the proposed Higgs factories, can further constrain the mixings down to $\mathcal{O} (10^{-2})$~\cite{Draper:2018ljh}. 

From Ref.~\cite{Kozaczuk:2019pet}, adopting the two-step transitions,
in order for SFOEWPT to happen in the small mixing limits, 
several conditions on the parameters of the scalar potential must be satisfied, such as
\be\begin{split}
a_2 &\gtrsim \frac{m_{h_1}^2}{4v^2} \frac{\Delta}{1-\Delta},\\
|b_3| &> \sqrt{\frac{9}{4}b_4 (2 m_{h_1}^2 - a_2 v^2 + 2T_{\text{EW}}^2 \beta)}, \\
b_4 &\gtrsim \frac{m_{h_1}^4 \Delta}{4 \lambda v^4 (1-\Delta)},
\end{split}\ee
where $T_{\text{EW}} \approx 140$ GeV, $\beta \equiv \frac{1}{12}(2 a_2 + 3 b_4)$ and
$\Delta \gtrsim 0.6 - 0.8$ empirically, while we take $\Delta = 0.7$ as in Ref.~\cite{Kozaczuk:2019pet}. This can be understood as we need strong couplings between the $S$ and $H$ to switch the EWPT from cross-over to strongly fist-order, and $a_2$ is the only such coupling in the $Z_2$ limits, therefore has lower bound.

Adopting these conditions, we randomly generate the data points in the following range
\be\begin{split}
& M_{h_1} \in [1,60]~\text{GeV}, 
\theta \in [10^{-10}, 10^{-2}], \\
& a_2 \in [\frac{m_{h_1}^2}{4v^2} \frac{\Delta}{1-\Delta}, 4 \pi],\\
& b_3 \in [\sqrt{\frac{9}{4}b_4 (2 m_{h_1}^2 - a_2 v^2 + 2T_{\text{EW}}^2 \beta)}, 4 \pi v],\\
& b_4 \in [\frac{m_{h_1}^4 \Delta}{4 \lambda v^4 (1-\Delta)}, 4\pi/3].
\label{eq:sample}
\end{split}\ee
After requring the data points to satisfy the SM constraints, we further use the {\tt Python} package {\tt CosmoTransitions}~\cite{Wainwright:2011kj} to solve Eq.~\ref{FOPT} to get $T_n$. Around $2\%$ of the data points can trigger a SFOEWPT. We realised that in our Eq.~\ref{VT}, only $\mathcal{O} (T^2)$ terms are considered, whereas leading $\mathcal{O} (T^3)$ terms from the gauge bosons are taken in Ref.~\cite{Kozaczuk:2019pet}\footnote{The effects of $\mathcal{O} (T^4)$ terms are considered in Ref.~\cite{Croon:2020cgk, Schicho:2022wty, Schicho:2021gca, Niemi:2021qvp, Ekstedt:2022bff}.}. However, the $\mathcal{O} (T^3)$ terms do not has appreciable effects, and  
our generated samples roughly match to the results in Ref.~\cite{Kozaczuk:2019pet}.

\subsection{Gravitational waves}

Stochastic GWs can be generated by a first-order EWPT. There are three sources which can contribute to the GWs spectrum including the bubble collisions, sound waves in plasma and the magneto-hydrodynamics turbulence~\cite{Mazumdar:2018dfl}. As discussed in Ref.~\cite{Ellis:2018mja}, the contribution from the bubble collisions is negligible due to the tiny energy transfer efficiency. Therefore, we only consider the latter two sources with the numerical expressions described in Ref.~\cite{Caprini:2015zlo, Espinosa:2010hh} \footnote{We also consider the suppression from the short duration of the sound wave period as discussed in Ref.~\cite{Ellis:2018mja, Guo:2020grp}.}. These expressions mainly depends on three input parameters, the transition latent heat over the radiation energy $\alpha$, the Universe expansion time scale over the phase transition duration $\beta/H$, and the bubble expansion velocity $v_b$. Larger $\alpha$~(larger energy released) and smaller $\beta/H$~(faster transition) can lead to stronger GWs. 
We take $v_b =$ 0.8 as a benchmark, while the former two parameters can be derived from the SFOEWPT pattern,
\be
\alpha=\frac{1}{g_*\pi^2T_n^4/30}\left(T\frac{\partial\Delta V_T}{\partial T}-\Delta V_T\right)\Big|_{T_n};\quad \beta/H=T_n\frac{d(S_{3}/T)}{dT}\Big|_{T_n},
\ee
where 
$\Delta V_T=V_T|_{T_n,(v^f,v_s^f)}-V_T|_{T_n,(0,v_s^i)}$ is effective potential difference between the true and false vacua, and $g_*\sim100$ is the number of relativistic degrees of freedom.

The GWs can be detected by the next-generation space-based interferometers, such as the LISA. The detectability can be characterised by the signal-to-noise ratio~(SNR), expressed as
\be
{\rm SNR}=\sqrt{\mathcal{T}\int_{f_{\rm min}}^{f_{\rm max}} df\left(\frac{\Omega_{\rm GW}(f)}{\Omega_{\rm LISA}(f)}\right)^2},
\ee
where $\Omega_{\rm LISA}$ is the sensitivity curve of the LISA detector~\cite{Caprini:2015zlo}, and $\mathcal{T}\approx 1.26 \times 10^8$~s = 4 years is the data-taking time. In the rest of the paper, we take SNR $> 10$ as the detection threshold for a six-link configuration LISA.

\section{Long-lived light scalars at the HL-LHC}
\label{sec:LLP}
The SFOEWPT can also be probed at colliders. It can lead to signals of resonantly produced scalars and correction to the Higgs couplings. As already mentioned in the introduction, many existing studies have focused on the cases with heavy scalars, while we focus on a light singlet scalar, enabling the exotic Higgs decay $h_2 \rightarrow h_1 h_1$ to happen. Such light singlet scalars with sufficiently small mixings, can be regarded as a LLP candidate, therefore the SFOEWPT can be tested at the lifetime frontiers of the LHC. In this section, we introduce the production cross section and decay length of the light scalar, and the relevant LLP analyses at the HL-LHC.

\subsection{Production and decay length of the light scalar}

\begin{figure}[htbp]
\centering
  \includegraphics[width=8cm, height=6cm]{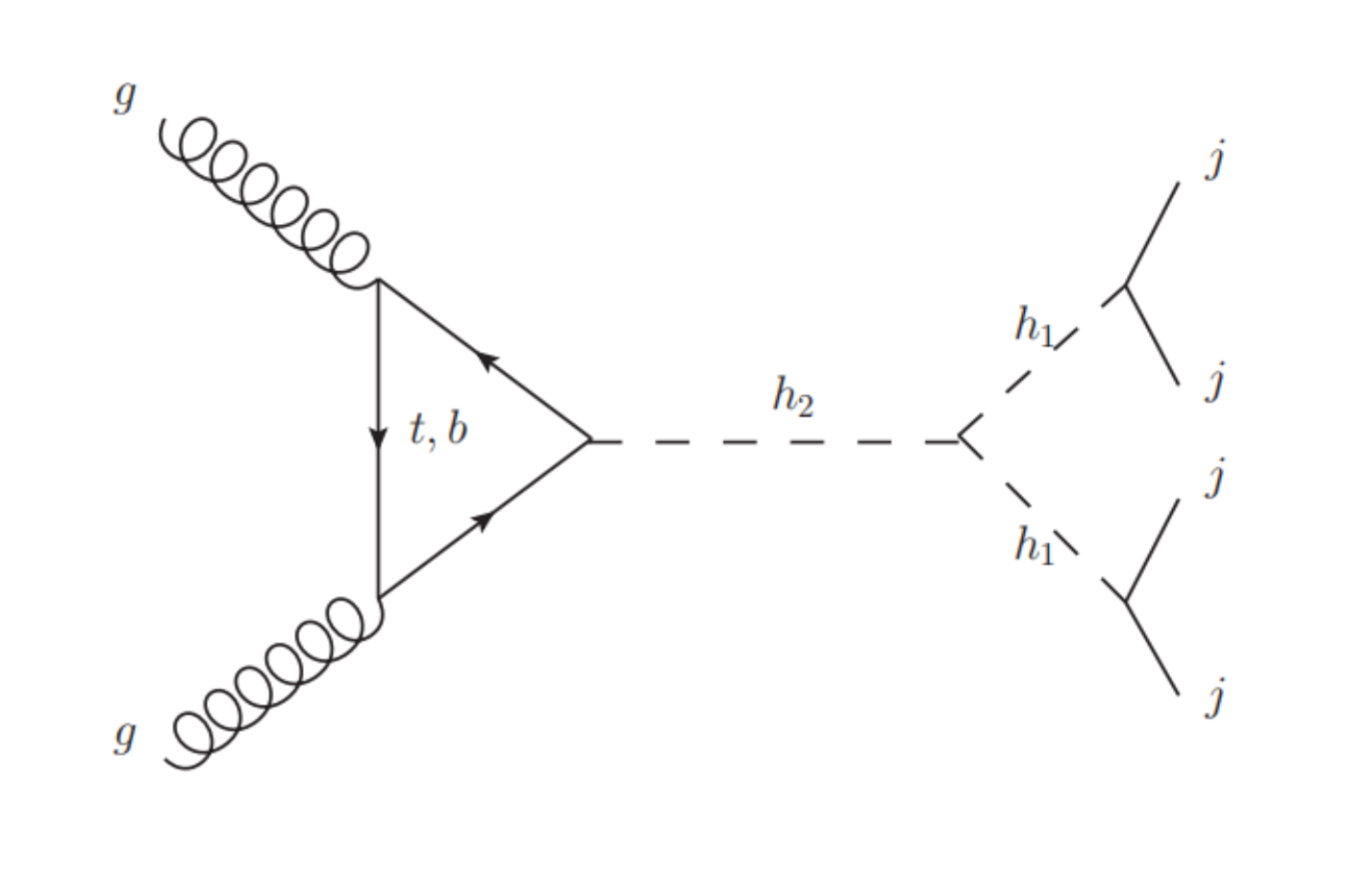}
  \caption{The Feynman diagram of the process $pp~(g g) \rightarrow h_2 \rightarrow h_1 h_1 \rightarrow j j j j$.}
\label{fig:feynman}
\end{figure}

\begin{figure}[htbp]
\centering
  \includegraphics[width=6.5cm, height=6.5cm]{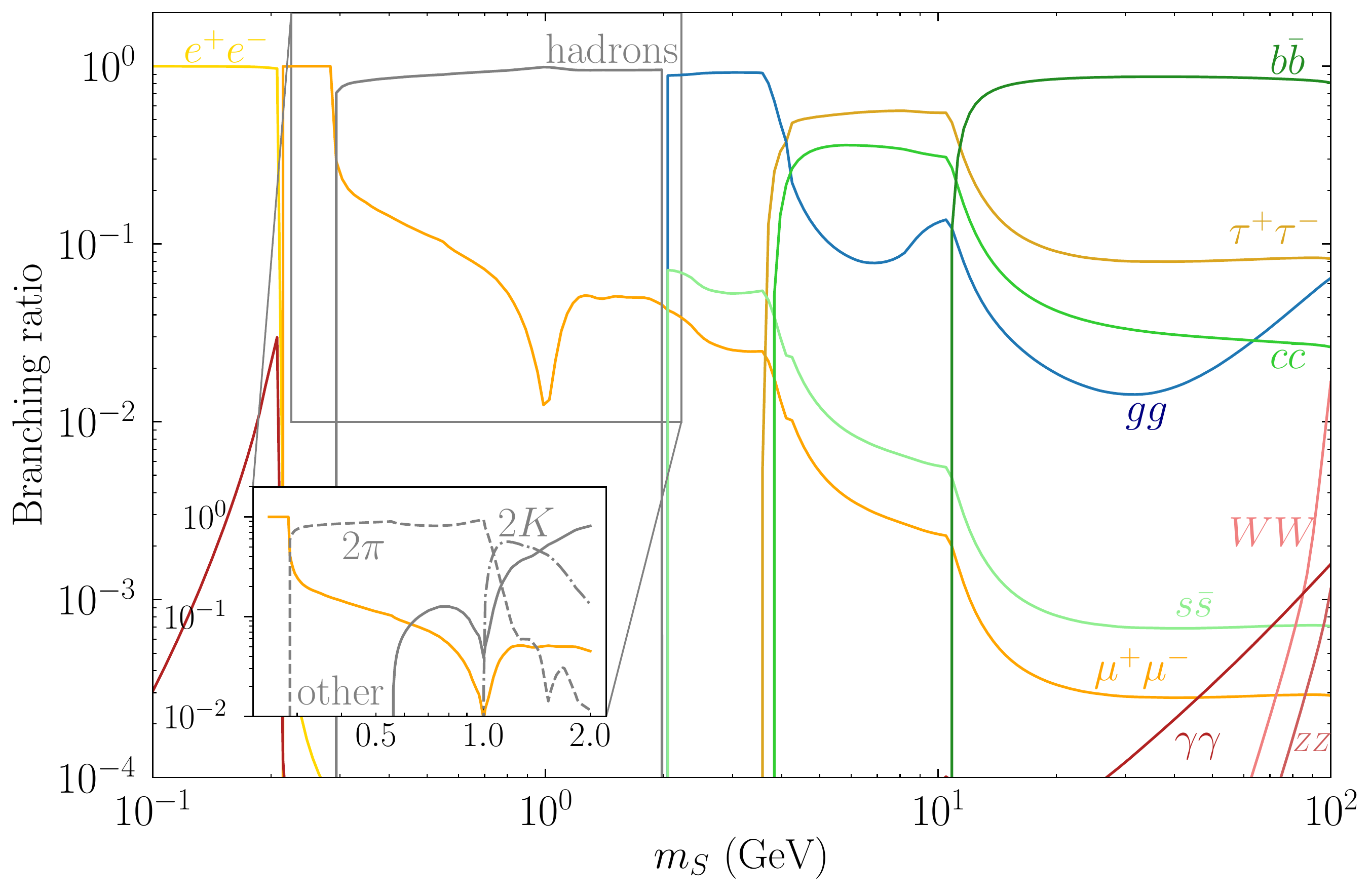}
  \includegraphics[scale=0.45]{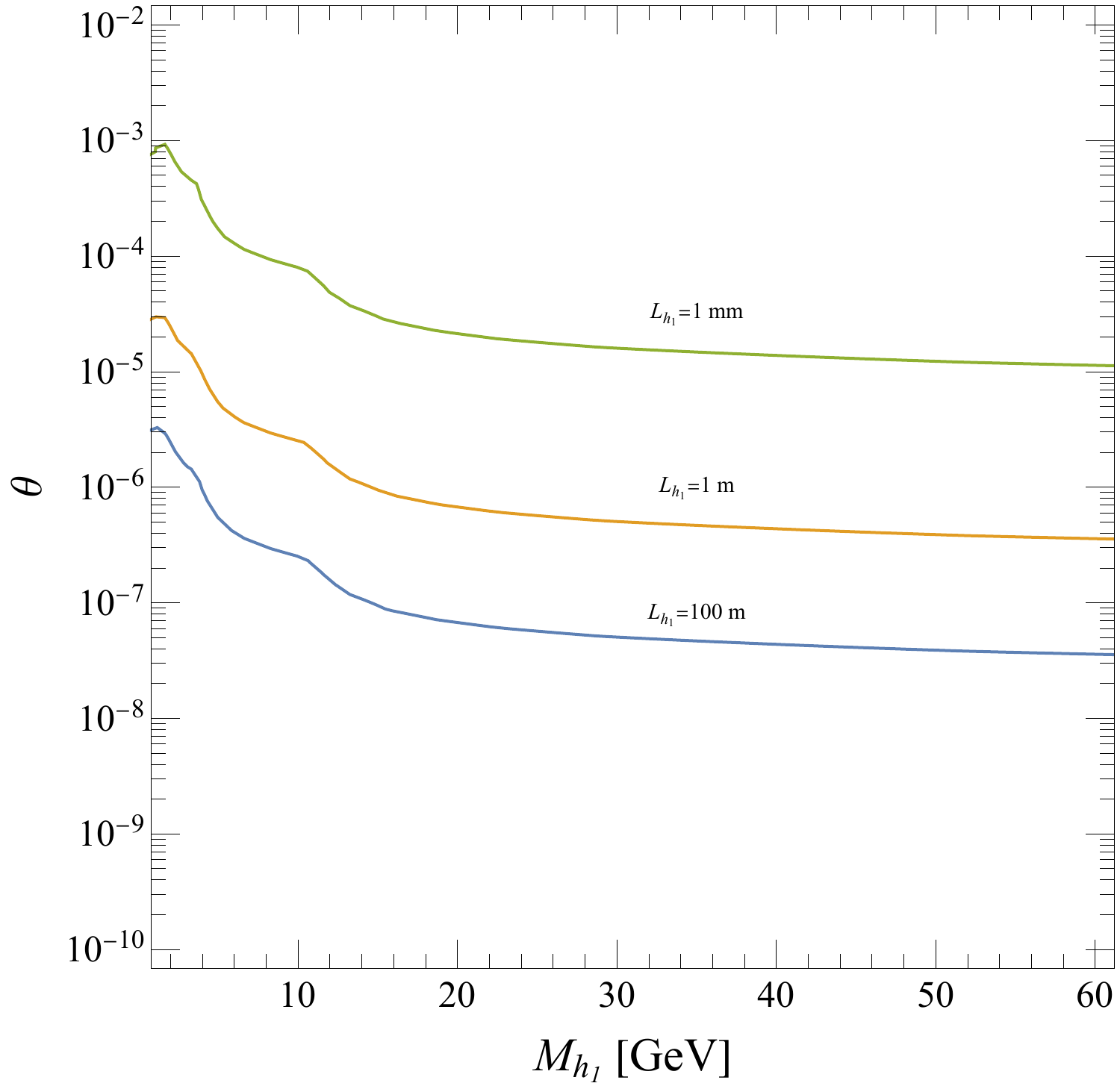}
  \caption{Left Panel: Estimated branching ratios of the scalar $h_1$ from Ref.~\cite{Gershtein:2020mwi}, where $m_s$ should be replaced by $M_{h_1}$. Right panel: The proper decay length of the $h_1$ calculated by Monte-Carlo simulation.}
\label{fig:scalar}
\end{figure}

Once kinematically allowed, the light singlet scalar $h_1$ can be pair-produced by the exotic decays of the SM Higgs at colliders. We have shown the corresponding Feynman diagram with $h_1 \rightarrow j j$ leading to 4 jets final states in Fig.~\ref{fig:feynman}. The cross section of the process is 
\be
\sigma_{pp \rightarrow h_2 \rightarrow h_1 h_1 \rightarrow j j j j }=c_\theta^2\times\sigma^{\text{SM}}_{h_2} \times \text{Br}_{h_2 \rightarrow h_1 h_1} \times \text{Br}^2_{h_1 \rightarrow j j} \approx \sigma^{\text{SM}}_{h_2} \times \text{Br}_{h_2 \rightarrow h_1 h_1} \times \text{Br}^2_{h_1 \rightarrow j j},
\ee
and 
\be\label{brh2h1h1}
\text{Br}_{h_2 \rightarrow X X} \approx \frac{\Gamma^{\text{SM}}_{h_2 \rightarrow XX}}{\Gamma^{\text{SM}}_{h_2} +\Gamma_{h_2\to h_1 h_1}},
\text{Br}_{h_2 \rightarrow h_1 h_1} \approx \frac{\Gamma_{h_2 \rightarrow h_1 h_1}}{\Gamma^{\text{SM}}_{h_2} +\Gamma_{h_2\to h_1 h_1}},
\ee
where $\Gamma^{\text{SM}}_{h_2} \approx 4$ MeV.
The partial decay
\be\label{h2h1h1}
\Gamma_{h_2\to h_1 h_1}=\frac{\lambda_{h_2 h_1 h_1}^2 }{32 \pi M_{h_2}}\sqrt{1-\frac{4M^2_{h_1}}{M^2_{h_2}}},
\ee
and the $h_2h_1 h_1$ coupling is defined by
\be
V \supset \frac{1}{2!}\lambda_{h_2 h_1 h_1}h_2h_1^2.
\ee
At the tree level, in the low mixing limit~\cite{Kozaczuk:2019pet},
\be\label{lambdah2h1h1}
\lambda_{h_2 h_1 h_1}= \left(\frac{1}{2} a_1 + a_2 v_s\right) s_\theta^3 + (2 a_2 v-6 \lambda v) c_\theta s_\theta^2
+\left(6 b_4 v_s + 2 b_3- 2 a_2 v_s - a_1\right) c_\theta^2 s_\theta- a_2 v c_\theta^3 \approx - a_2 v,
\ee
hence $\text{Br}_{h_2 \rightarrow h_1 h_1}$ is only controlled by $a_2$.
Determining $\text{Br}_{h_1 \rightarrow j j}$ involves calculations of hadronic effects, as discussed in Ref.~\cite{Gershtein:2020mwi}, and is shown in Fig.~\ref{fig:scalar}~left. For the $M_{h_1}$ of our interests, $\text{Br}_{h_1 \rightarrow j j} \approx 100~\%$ except the window where $h_1 \rightarrow \tau \tau$ just opens for 5 GeV $ \lesssim M_{h_1} \lesssim$ 10 GeV.
In general, the partial decay width of the $h_1$,
\be
\Gamma_{h_1 \rightarrow XX}=s_\theta^2 \times \Gamma^{\text{SM}}_{h_1 \rightarrow XX} \sim 0,
\ee
where $\Gamma^{\text{SM}}_{h_1 \rightarrow XX}$ is the partial decay width with the SM couplings. If the mixing is tiny, the  $h_1$ can become a LLP.

To simulate the signal events which can be used for analyses of LLP searches, we use the Monte-Carlo generator {\tt MadGraph5aMC$@$NLO} v2.6.7~\cite{Alwall:2014hca} with input of a Universal FeynRules Output (UFO)~\cite{Degrande:2011ua} file generated by {\tt FeynRules}~\cite{Alloul:2013bka,Christensen:2008py} based on a scalar singlet model. The initial and final state parton shower, hadronization, heavy hadron decays, etc are handled by {\tt PYTHIA} v8.235~\cite{Sjostrand:2014zea}. The clustering of the events is performed by {\tt FastJet} v3.2.1~\cite{Cacciari:2011ma}. Detector level simulation is not considered at the current stage, while geometrical acceptance of the detectors are roughly simulated by the analyses in the following subsection.

With the above setup, we show the proper decay length of the $h_1$ calculated by Monte-Carlo simulation in Fig.~\ref{fig:scalar}~right. We verify that our results roughly match to Ref.~\cite{Gershtein:2020mwi}, which only has $\mathcal{O}(1)$ difference to Ref.~\cite{Winkler:2018qyg}. For $10^{-7} \lesssim \theta \lesssim 10^{-3}$, the $h_1$ decays meters away from the interaction point, can be regarded as a LLP candidate. In the next subsection, we introduce several analyses at the HL-LHC for the searches of such LLPs at far detectors including the FASER and MAPP, as well as the CMS-Timing.

\subsection{LLP searches at the HL-LHC}
Aiming at probing LLPs, several specialized far detectors are proposed including the FASER and MAPP, as well as the CODEX-b~\cite{Gligorov:2017nwh}, FACET~\cite{Cerci:2021nlb} and MATHUSLA~\cite{Chou:2016lxi}. The upgrade of the CMS detector also include a timing detector which can be used for the same purpose.
Following them, a substantial number of analyses have been put forward. We only focus on the detectors which are already in installation, and adopt the analyses from Ref.~\cite{Berlin:2018jbm, Liu:2022ugx}. These analyses have already been introduced in details in the literature, a summary of them is provided in Table~\ref{tab:LLP}. The signal of LLPs is defined such the particle is decayed inside the detector volume, and the final states can pass the trigger. We use the phase 2 design of the FASER to maximize the geometrical acceptance. The CMS-Timing analysis requires one jet from initial state radiation~(ISR) to timestamp the event, which we denote as $j_i$. The signal event is identified if the jets from the LLP decays are sufficiently delayed comparing to the ISR jet. And the time delay can be estimated as $\Delta t \approx L_{h_1}/ \beta_{h_1} + L_{j} / c - L_{j_i} / c$~\cite{Liu:2018wte,Berlin:2018jbm}.  

\begin{table}
	\centering
	\begin{tabular}{|l|c|c|c|c|c|c|}
		\hline
		Detectors & $L_x$~[m]  & $L_y$~[m] & $L_{xy}$~[m] & $L_z$~[m] & trigger\\
		\hline
		FASER  & $-$  & $-$  & $[0,1]$ & $[475, 480]$ &  $E_{\text{vis}} > 100~\text{GeV}$ \\
		\hline
		MAPP  & $[3,6]$  & $[-2,1]$  & $-$ & $[48, 61]$ &  $E_{\text{track}} > 0.6~\text{GeV}$ \\
		\hline
		CMS-Timing  & $-$  & $-$  & $[0.05, 1.17]$ & $[-3.04, 3.04]$ &  $p_T(j_i, j) >$ 30 GeV, $\Delta t > 0.3~\text{ns}$ \\
		\hline
	\end{tabular}
	\caption{The LLP analyses corresponding for the FASER~\cite{Feng:2017uoz}, MAPP~\cite{Pinfold:2019nqj} and CMS-Timing~\cite{Butler:2019rpu, Liu:2018wte} detectors. $p_T(j_i)$ is the transverse momentum of the jet from initial state radiation, $\Delta t$ is the time delay between the jets from initial state radiation and the LLP decays.  }
	\label{tab:LLP}
\end{table}

After applying these analyses, the overall signal events can be expressed as,
\begin{align}
N_{signal} &= \sigma_{pp \rightarrow h_2} (M_{h_1}, s_\theta) \times L \times \text{Br}_{h_2 \rightarrow h_1 h_1} (a_2, M_{h_1}) \times \text{Br}_{h_1 \rightarrow j j}^2(M_{h_1}) \\ \nonumber
&\times \epsilon_{kin}(M_{h_1}) \times \epsilon_{geo}(M_{h_1}, s_\theta), 
\end{align}
where $L$ is the integrated luminosity, $\epsilon_{kin, geo}$ are the efficiencies due to trigger requirements and geometrical acceptance, respectively. The reconstruction efficiency is not considered so far, assumed to be 100 \%. 

\begin{figure}[htbp]
\centering
  \includegraphics[scale=0.42]{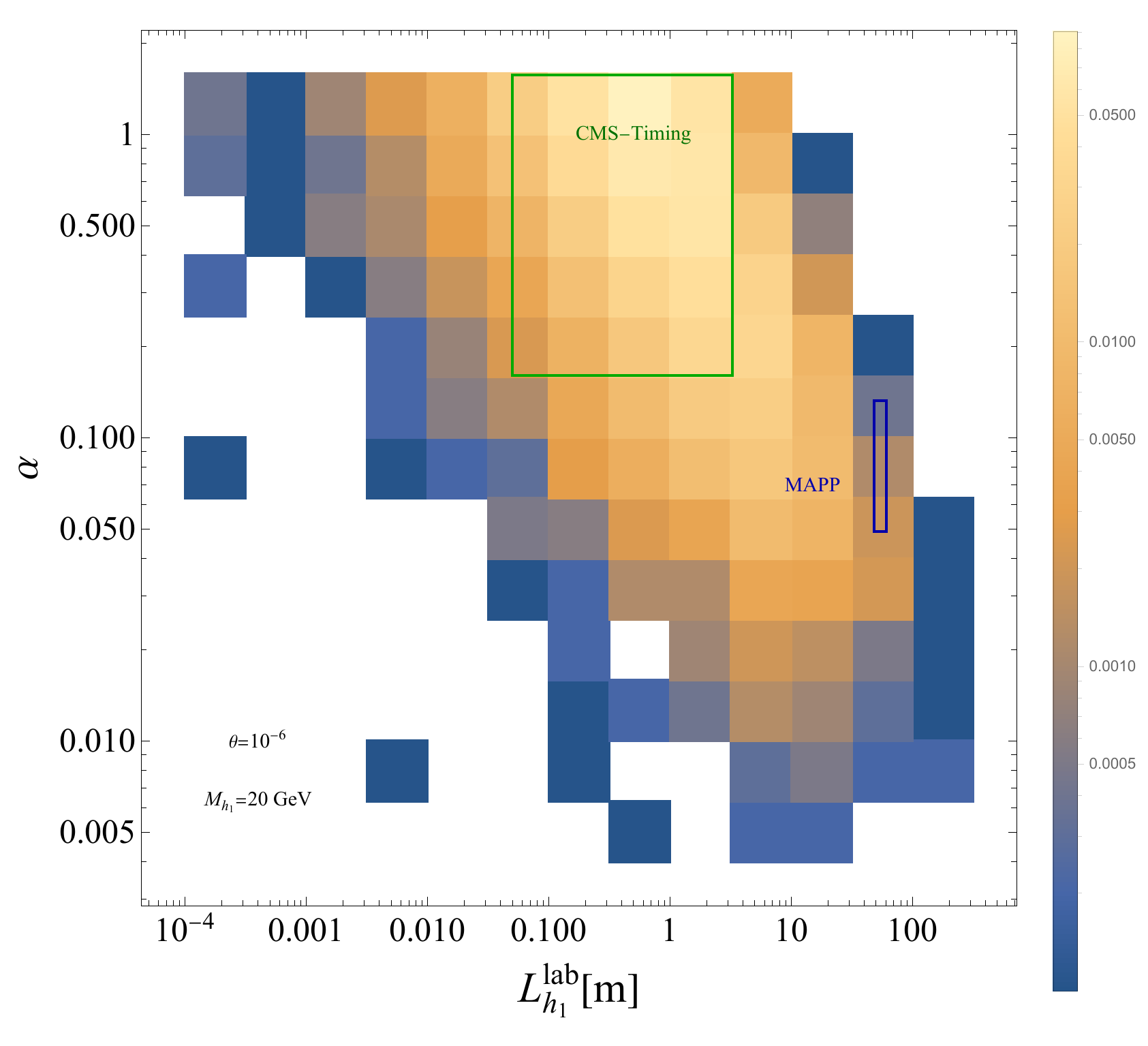}
    \includegraphics[scale=0.42]{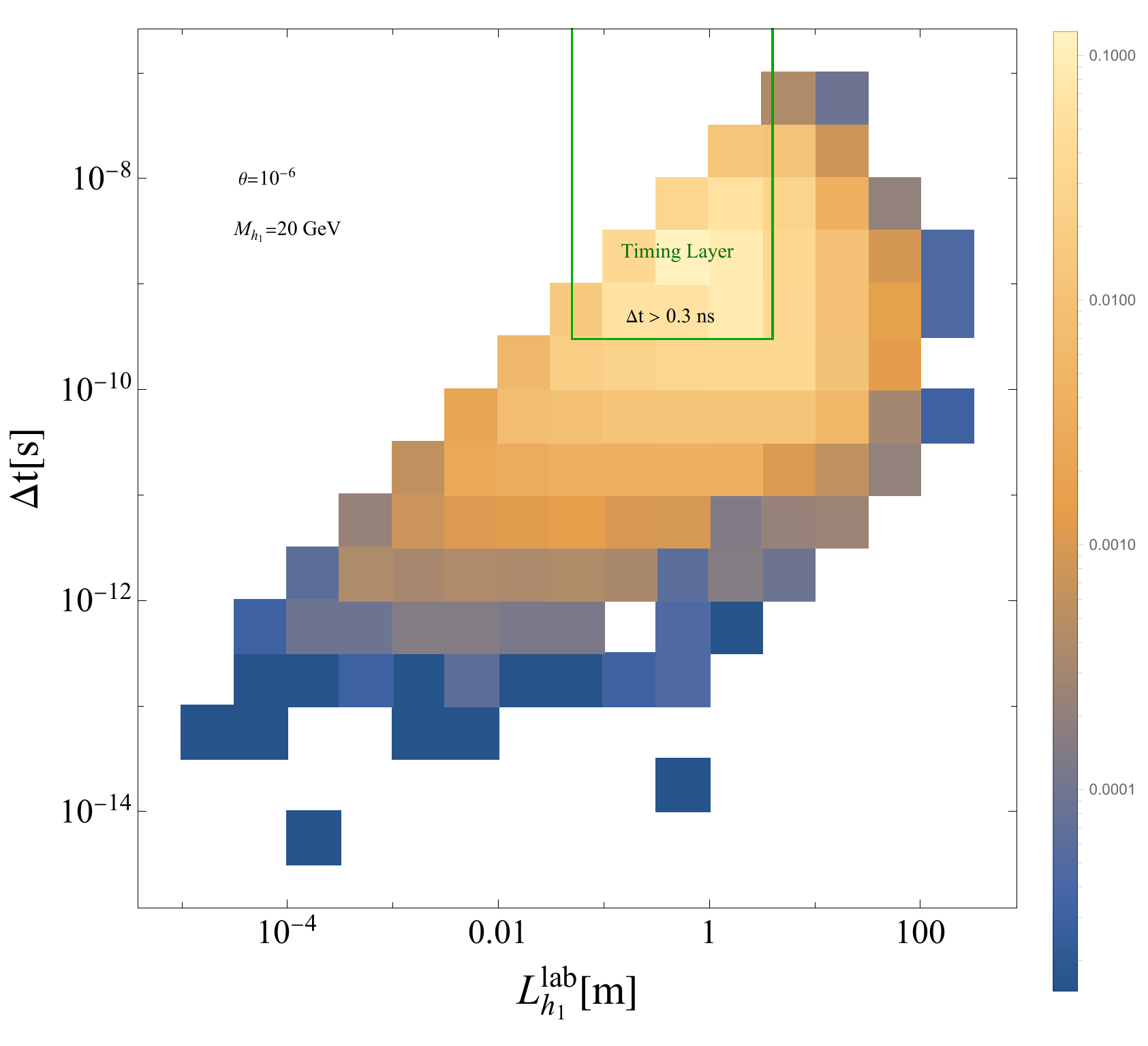}
  \caption{Left panel: ($L_{h_1}^{\text{lab}}$, $\alpha$) distribution. Right panel: ($L_{h_1}^{\text{lab}} $, $\Delta t$) distribution. The probability distribution density is expressed by the colors as indicated by the bars at the right handed side of each panel.}
  \label{fig:dis}
\end{figure}

The effects of $\epsilon_{ geo}$ is prominent, varies at different LLP detectors, depend on the angle between the LLPs and the beam line $\alpha$, the time delay $\Delta t$, and
the lab decay length of the LLP $L_{h_1}^{\text{lab}} $. We therefore display the distribution of these variables for the $h_1$ in Fig.~\ref{fig:dis}. The probability distribution density is expressed by the colors as indicated by the bars at the right handed side of each panel.
In the left panel, we show the $\alpha$ and $L_{h_1}^{\text{lab}} $ distribution for the LLP at a benchmark where $M_{h_1} =$ 20 GeV and $\sin \theta = 10^{-6}$. The coverage of the CMS-Timing and MAPP are also shown for comparison. The CMS-Timing covers the region where the LLPs are highly distributed,
which is much larger than the one from the MAPP, hence expected to yield higher $\epsilon_{geo}$. In spite of that, the FASER has failed to reach any coverage at all.
Since the $\epsilon_{ geo}$ of the CMS-Timing also depends on the $\Delta t$, we show the distribution of the $\Delta t$ and $L_{h_1}^{\text{lab}} $ in the right panel for the same benchmark. Most of the LLPs follow the line where $L_{h_1}^{\text{lab}} \sim c \Delta t$. And the CMS-Timing detector has shown substantial coverage as well.

\begin{figure}[htbp]
\centering
  \includegraphics[scale=0.45]{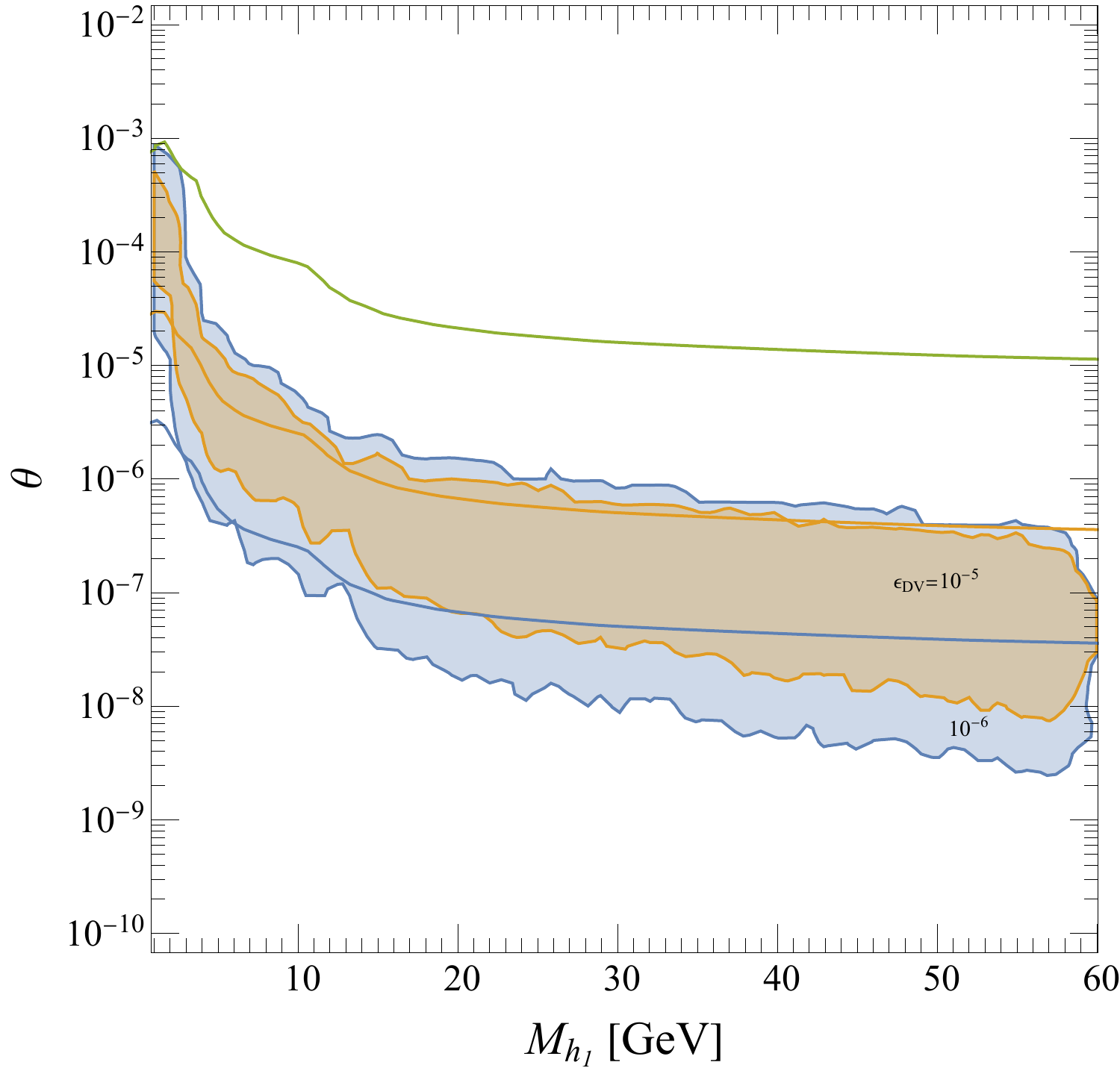}
    \includegraphics[scale=0.45]{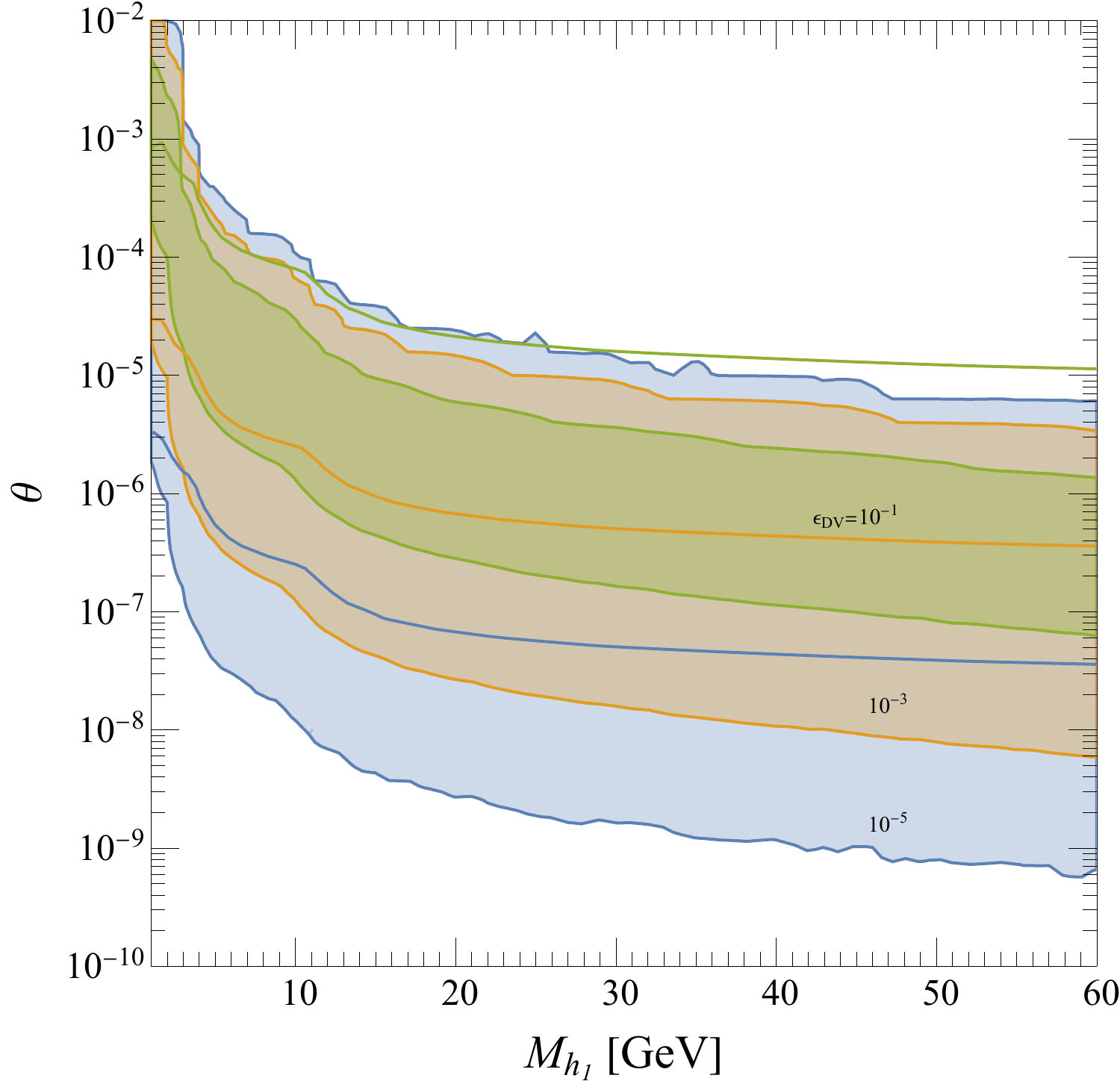}
  \caption{Left panel: Detection efficiency of the MAPP as a function of $M_{h_1}$ and $\theta$. Right panel: Same but for CMS-Timing.}
  \label{fig:eff}
\end{figure}

The overall efficiencies $\epsilon_{DV} \equiv \epsilon_{kin} \times \epsilon_{geo}$ for the two detectors, CMS-Timing and MAPP, as a function of $M_{h_1}$ and $\theta$
are shown in Fig.~\ref{fig:eff} left and right, respectively. $\epsilon_{DV}$ only depends on the decay length and masses of the $h_1$, therefore do not depend on the production channels, can be used to determine the sensitivity for other channels as well. The efficiencies of the MAPP and CMS-Timing peak at where $L_{h_1} \sim 1-100$~m. As expected, $\epsilon_{DV}$ of the CMS-Timing can reach as high as $10^{-1}$, which is about four magnitude larger than the one from the MAPP.

\section{Sensitivity}
\label{sec:sen}
Since we already obtain the number of the signal events above, now we need to estimate the background to derive the sensitivity. In general, the background for the LLPs mainly consists of the long-lived mesons decays in the SM, real particles produced via interactions with the detector, real particles originating from outside the detector, fake-particle signatures as well as the algorithmically induced fakes~\cite{Alimena:2019zri}. The first one can be cut away by the mass fit, while the other ones are hard to model but still suppressed. Given our selected analyses, the background events can be regarded as negligible~\cite{Liu:2018wte,Frank:2019pgk}. Therefore, we adopt a Poisson distribution, and
only require \be
N_{signal} =3.09,
\ee
at 95 \% confidence level~(C.L.) to obtain sensitivity at certain parameter points~\cite{ParticleDataGroup:2020ssz}.

The SFOEWPT parameter space can be projected into the plane ($M_{h_1}$, $\text{Br}_{h_2 \rightarrow h_1 h_1}$), while the  $N_{signal}$ is additionally dependent on the $\theta$, since it alters the decay length. In order to show this dependency, we selected three benchmark scenarios as defined below,
\begin{itemize}
    \item Scenario A, fixed $\theta = 10^{-4}$.
    \item Scenario B, fixed $\theta = 10^{-6}$.
    \item Scenario C, running $\theta \in [10^{-10},10^{-2}]$.
\end{itemize}
The other parameters are chosen as described in Eq.~\ref{eq:sample}.

\begin{figure}[htbp]
\centering
  \includegraphics[scale=0.43]{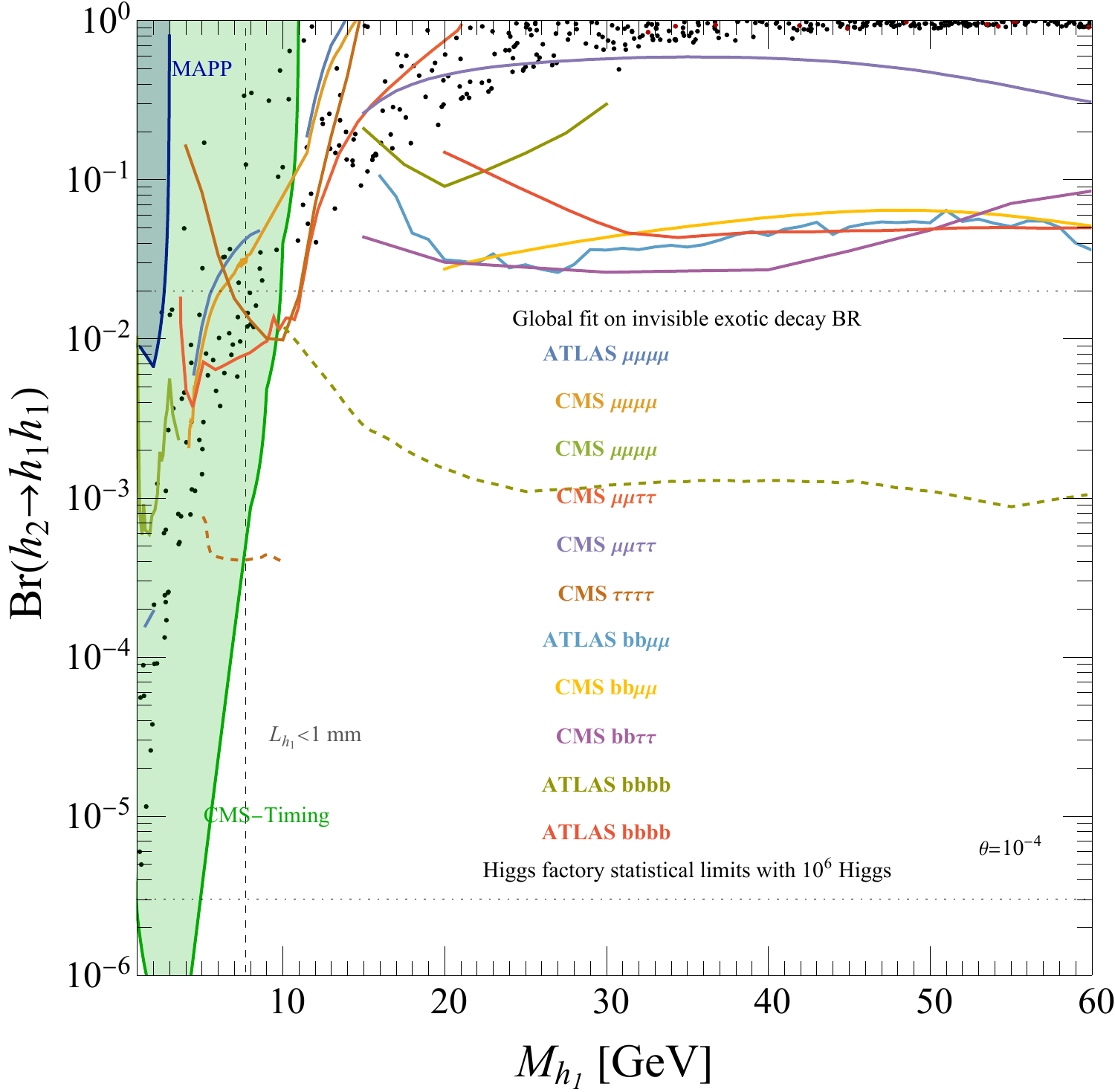}
    \includegraphics[scale=0.43]{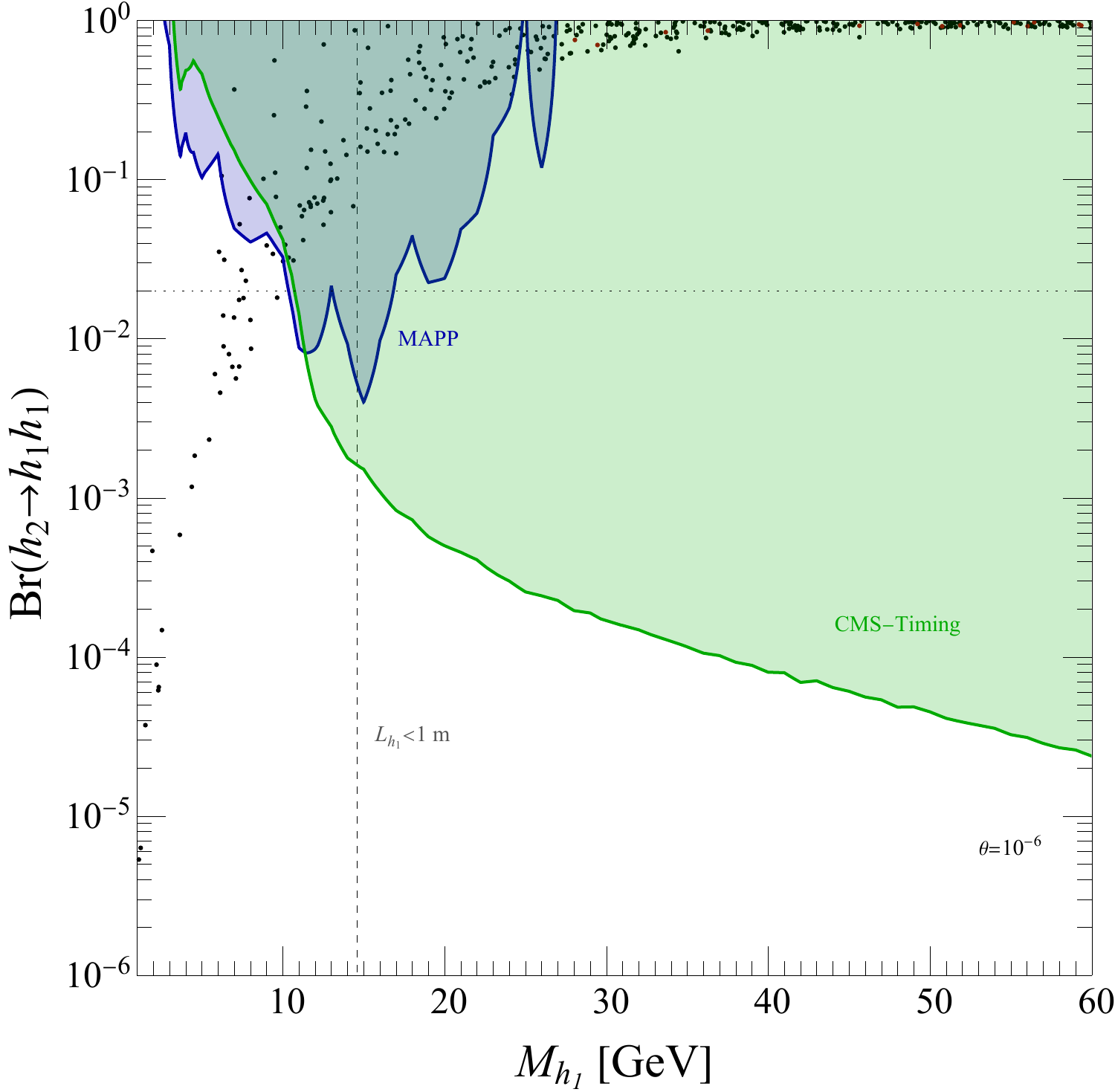}
  \caption{Left: The expected probe limits on $\text{Br}_{h_2 \rightarrow h_1 h_1}$ from different LLPs analyses for Scenario A. The scatter points are the SFOEWPT data, in which black, red colors represent ${\rm SNR}\in[0,10)$, $[10,\infty)$ for the detectability of the GWs at the LISA, respectively. The projected sensitivity of the exotic Higgs decays from prompt final states at the HL-LHC~(solid curves) and Higgs factories~(dashed curves) are also shown for comparison~\cite{Carena:2022yvx,Gershtein:2020mwi}.  The upper and lower horizontal dotted lines are the
expected upper limit for invisible Higgs exotic decay branching ratio at the HL-LHC (2\%~\cite{deBlas:2019rxi}) and
statistical limit of $10^6$ Higgs at future lepton colliders~\cite{Carena:2022yvx}, respectively. Right: Same but for Scenario B.}
  \label{fig:fixsa}
\end{figure}

The resulting sensitivity for the SFOEWPT is shown in Fig.~\ref{fig:fixsa} for Scenario A and B. For Scenario A, only $M_{h_1} \lesssim $ 10 GeV can lead to LLPs, as pointed out by the vertical dashed black line. Due to the high detection efficiency as shown in Fig.~\ref{fig:eff}, the CMS-Timing can reach $\text{Br}_{h_2 \rightarrow h_1 h_1} \sim 10^{-6}$, which is four magnitude better than the MAPP. Hence, it can cover almost all the SFOEWPT points when $M_{h_1} \lesssim $ 10 GeV, where the promptly exotic Higgs decays at the HL-LHC~(solid curves) and Higgs factories~(dashed curves)~\cite{Carena:2022yvx} can not probe since the final states are no-longer prompt. However, it failed to reach any points for $M_{h_1} \gtrsim $ 10 GeV where the prompt searches come into rescue, and probe most of the parameter points there. The sensitivity from the prompt analyses is still shown for $M_{h_1} \lesssim $ 10 GeV, but the validity of it needs to be clarified. The limits on $\text{Br}_{h_2 \rightarrow h_1 h_1}$ should be re-scaled since the detection efficiency drops when the $h_1$ has sufficient large decay length, i.e. $\epsilon_{\text{prompt}} \approx 1- e^{-\text{1mm}/L^{\text{lab}}_{h_1}}$. Considering a average Lorentz factor of $\mathcal{O}(10)$ from the boost of the $h_1$ so $L^{\text{lab}}_{h_1} \sim 10$~mm, $\epsilon_{\text{prompt}} \sim 10^{-1}$. Therefore, the prompt searches should only reach $\text{Br}_{h_2 \rightarrow h_1 h_1}\sim 10^{-2}$ at most for $M_{h_1} \lesssim $ 10 GeV.
Due to the low detection efficiency and the requirement on the masses of the $h_1$, the MAPP can merely probe any SFOEWPT points. As for the detectability of the GWs at the LISA, we use black, red colors of the SFOEWPT points to represent ${\rm SNR}\in[0,10)$, $[10,\infty)$, respectively.
Almost all the SFOEWPT points can not lead to sufficient ${\rm SNR}$ at the LISA, except a few points around $M_{h_1} \gtrsim $ 10 GeV, which are already covered by the prompt searches at the HL-LHC, and the expected upper limit for invisible exotic Higgs decay branching ratio at the HL-LHC (2\%)~\cite{deBlas:2019rxi}. Hence, combing the CMS-Timing and prompt searches, the collider seaches for the exotic Higgs decays can probe SFOEWPT parameter space remarkably, where the LISA can merely cover. 

As shown in the right panel of the Fig.~\ref{fig:fixsa}, the Scenario B does not have significant difference in the GWs detectability comparing to the Scenario A. Nonetheless, the sensitivity of the LLP analyses have shown drastic changes owing to the lower $\theta$. Now the LLP analyses require $M_{h_1} \gtrsim $ 10 GeV, since lighter $h_1$ escapes the detector volumes. The CMS-Timing and MAPP are now sensitive to $M_{h_1} \gtrsim $ 10 GeV, and reach $\text{Br}_{h_2 \rightarrow h_1 h_1} \sim 10^{-4}$ and $10^{-2}$, respectively. In this scenario, within the whole parameter space, the $h_1$ is long-lived enough, cannot be captured by the prompt searches, therefore only the LLPs analyses and the invisible Higgs exotic decay at the HL-LHC can probe the SFOEWPT.

\begin{figure}[htbp]
\centering
  \includegraphics[scale=0.5]{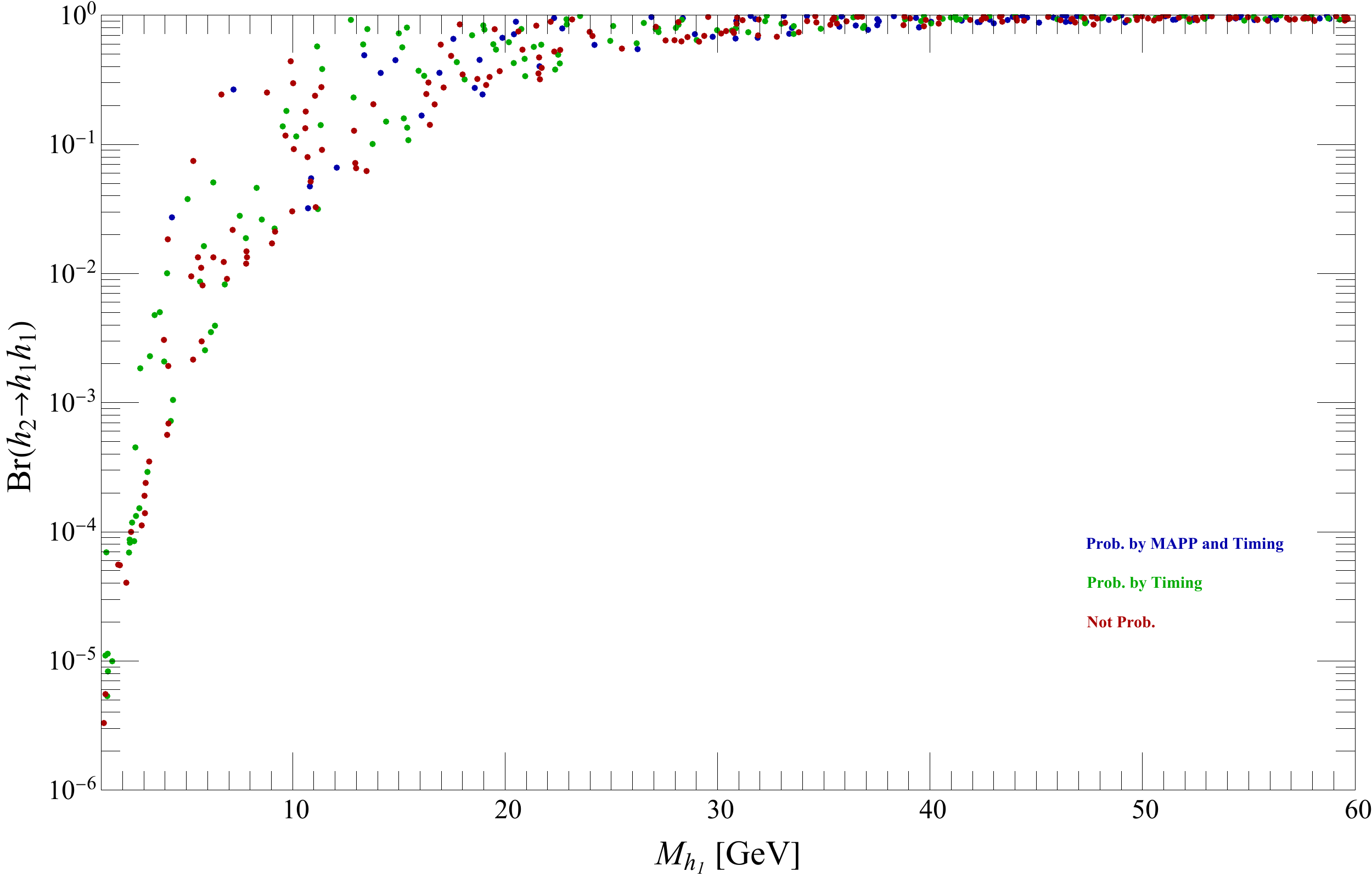}
  \caption{Same as Fig.~\ref{fig:fixsa}, but for 
  Scenario C. The blue, green and red colors of the scatter points represent whether they are probed by both the MAPP and CMS-Timing, only the CMS-Timing, or not any of them. }
  \label{fig:runsa}
\end{figure}

Fig.~\ref{fig:runsa} shows the sensitivity for the Scenario C. Since the $\theta$ varies, the detectability of the LLP analyses for the SFOEWPT points are different even at same $M_{h_1}$. Overall, the CMS-Timing have shown better sensitivity than the MAPP. 
Compared to the MAPP, the CMS-Timing can cover all of the SFOEWPT points which are probed by the MAPP, and probe an appreciable number of additional points especially for light $h_1$ with $M_{h_1} \lesssim $ 10 GeV. There are still a substantial number of the points which are not probed by any of the LLP analyses. The sensitivity from the prompt searches and invisible Higgs exotic decays are not shown here, as the decay length of the $h_1$ varies, so the final states might not be prompt or invisible. 

\section{Conclusion}
\label{sec:conclu}
The pattern of the electroweak phase transition is an important question of the particle physics, as it can reveal the early history of the Universe. To answer this question, we concentrate on the SFOEWPT triggered by a light scalar, and it can be probed by the exotic Higgs decays to a pair of the scalar $h_1$ at colliders. 

We generate the points for the SFOEWPT in this case. As already pointed out by Ref.~\cite{Carena:2019une,Kozaczuk:2019pet, Carena:2022yvx}, light $h_1$ can lead to sufficiently large $\text{Br}_{h_2 \rightarrow h_1 h_1}$ which can be probed by the searches for the promptly exotic Higgs decays at the HL-LHC and Higgs factories. In this article, we additionally show that for very light $h_1$, the scalar can be regarded as a LLP for certain Higgs mixings $\theta$, therefore can be probed by the existing LLP analyses based at the FASER, MAPP and CMS-Timing detectors. In such parameter space, the prompt analyses are not valid anymore, since the final states are no-longer prompt, LLP analyses are one of the only ways to test the SFOEWPT.
Thanks to the low background and high detection efficiencies, we show that both the LLP analyses of the MAPP and CMS-Timing probe very low $\text{Br}_{h_2 \rightarrow h_1 h_1}$ for certain $\theta$, while the FASER fail to reach any parameter space due to the small geometrical acceptance. This helps to probe substantial part of the SFOEWPT points, especially using the analysis at the CMS-Timing detector.

The possibility of using the GWs to probe the SFOEWPT is also considered. However, it is shown for light $h_1$, the GWs detector, LISA is not likely to observe sufficient signals. Even though a few of the SFOEWPT points can be probed at the LISA, they are already covered by the projected limits from promptly/invisible exotic Higgs decays at the HL-LHC.

We have shown that the LLP and prompt analyses are complementary in probing the SFOEWPT. The LLP analyses are in particular sensitive to the cases where the $M_{h_1}$ and $\theta$ can lead to the scalar with meters of decay length. While the prompt analyses can probe SFOEWPT points only if the $\theta$ is sufficiently large, so the scalar possess decay length less than millimeter.

\begin{acknowledgments}
We thank Ke-pan Xie for sharing the codes and useful discussions. This work is supported by the Natural Science Foundation of Jiangsu Province (Grants No.BK20190067). Wei Liu is supported by the 2021 Jiangsu Shuangchuang (Mass Innovation and Entrepreneurship) Talent Program (JSSCBS20210213). Hao Sun is supported by the National Natural Science Foundation of China (Grant No.12075043, No.12147205).
\end{acknowledgments}

\bibliographystyle{JHEP}
\bibliography{submit.bib}

\end{document}